\documentclass[a4paper,11pt]{article}
\usepackage{aaskaiid}
\usepackage{aas-macros}
\usepackage{graphicx}
\usepackage{orcidlink}

\newcommand\arcsec{\ensuremath{^{\prime\prime}}}

\newcommand{\cheung}[1]{{#1}}

\newcommand{\postreview}[1]{{#1}}
\newcommand{\postacceptance}[1]{{#1}}

\title{Probing the Physics of the Solar Corona with High-Resolution Observations using SKA-Mid}
\ShortTitle{High-resolution Solar Physics with SKA-Mid}

\author[1]{Mark~C.~M.~Cheung\,\orcidlink{0000-0003-2110-9753}}
\ShortName{Cheung et al.} 
\author[2]{John~S.~Morgan\,\orcidlink{0000-0001-9224-5483}}
\author[3]{Rohit Sharma\,\orcidlink{0000-0003-0485-7098}}
\author[4,5]{Sven Wedemeyer\,\orcidlink{0000-0002-5006-7540}}
\author[6,7]{Gregory Fleishman\,\orcidlink{0000-0001-5557-2100}}
\author[8]{Matthias Rempel\,\orcidlink{0000-0001-5850-3119}}

\affiliation[1]{CSIRO Space and Astronomy, P.O. Box 76, Epping, NSW 1710, Australia}
\emailAdd{mark.cheung@csiro.au}
\affiliation[2]{CSIRO Space and Astronomy, P.O. Box 1130, Bentley, WA 6102, Australia}
\emailAdd{john.morgan@csiro.au}
\affiliation[3]{Space, Planetary \& Astronomical Sciences \& Engineering (SPASE), Indian Institute of Technology Kanpur, Kalyanpur, Kanpur, 208016, Uttar Pradesh, India}
\emailAdd{rsharma@iitk.ac.in}
\affiliation[4]{Rosseland Centre for Solar Physics, University of Oslo, Oslo, Oslo, N-0315, Norway}
\affiliation[5]{Institute of Theoretical Astrophysics, University of Oslo, Oslo, Oslo, N-0315, Norway}
\emailAdd{sven.wedemeyer@astro.uio.no}
\affiliation[6]{Center For Solar-Terrestrial Research, New Jersey Institute of Technology, Newark, NJ 07102, USA}
\emailAdd{gfleishm@njit.edu}
\affiliation[7]{Institut f\"ur Sonnenphysik (KIS), Georges-K\"{o}hler-Allee 401 A, D-79110 Freiburg, Germany}
\affiliation[8]{High Altitude Observatory, NSF NCAR, P.O. Box 3000, Boulder, Colorado 80307, USA}
\emailAdd{rempel@ucar.edu}

\abstract{We examine the capability of the SKA-Mid telescope for probing solar coronal structure and evolution. Using radiative magnetohydrodynamics simulations of a solar flare, we synthesise radio emission in bands available to SKA-Mid. SKA-Mid observations would provide important constraints on plasma processes that are important to diverse astrophysical environments, including flares/eruptions, magnetic reconnection, particle acceleration, and gyro-emission processes. Observations with the AA4 configuration would \cheung{reveal fine-scale structures in the solar atmosphere,} making SKA-Mid a complementary telescope to other high resolution solar observatories operating at optical, infrared, and XUV wavelengths. \postreview{Estimates of coronal scattering suggest that angular broadening rather than instrumental resolution (0.1 \arcsec at 10 GHz) will limit the detection of the finest structure under typical conditions. The extent of angular broadening depends on the nature of turbulence in the corona, and the longer baselines of SKA-Mid will serve to provide new observational constraints on coronal structure.}}

\begin{document}
\maketitle

\section{Introduction}
\label{sec:intro}
The solar atmosphere is a natural laboratory for studying fundamental plasma physics processes that are expected to operate in diverse astrophysical environments. These processes include MHD turbulence, magnetic reconnection, magnetically-driven flares and eruptions, magnetoacoustic waves and instabilities, astrophysical shocks, particle acceleration and much more. In the past decade, a considerable number of scientific discoveries and advances in our understanding of solar physics have been enabled by observations at arcsecond and subarcsecond resolutions (corresponding to $L \lesssim$720\,km on the Sun), particularly at microwave, optical, UV, and XUV wavelengths \citep[e.g.,][]{Winebarger2013,Tian2014,Hinode:2019,2020NatAs...4.1140C,IRIS:2021,2020Sci...367..278F,2022Natur.606..674F,Harra:2025}. The highly intermittent and structured dynamics in the solar atmosphere at these scales have raised new questions about the physical mechanisms underlying coronal heating~\citep{DePontieu:MUSE2022} and flares/eruptions~\citep{Cheung:MUSE2022}. 

\citet{Gary:2023} presents a thorough review of recent solar science enabled by the current generation of radio astronomy facilities such as the Expanded Owens Valley Array~\citep[EOVSA;][]{Gary2018}, the Murchison Widefield Array~\citep[MWA;][]{Tingay:MWA}, the Jansky Very Large Array (JVLA), and the Atacama Large Millimeter Array~\citep{ALMA2009}.  For example, millimeter observations with ALMA~\citep{ALMA2009} have enabled a limited number of studies of the thermal structure of the solar chromosphere~\citep{Wedemeyer2016, Bastian2018} at close to arcsecond resolution. \postacceptance{Under construction, the Square Kilometre Array Observatory (SKAO) will have wide frequency coverage (50 MHz - 15.4 GHz), which will enable a range of solar science cases summarised in the overview chapter \citep{Zucca01.2026.SKA}}.


\section{SKA-Mid as a High Resolution Solar Observatory}\label{sec:SKAmidSolar}
\begin{figure}
\centering
\includegraphics[width=0.46\textwidth]{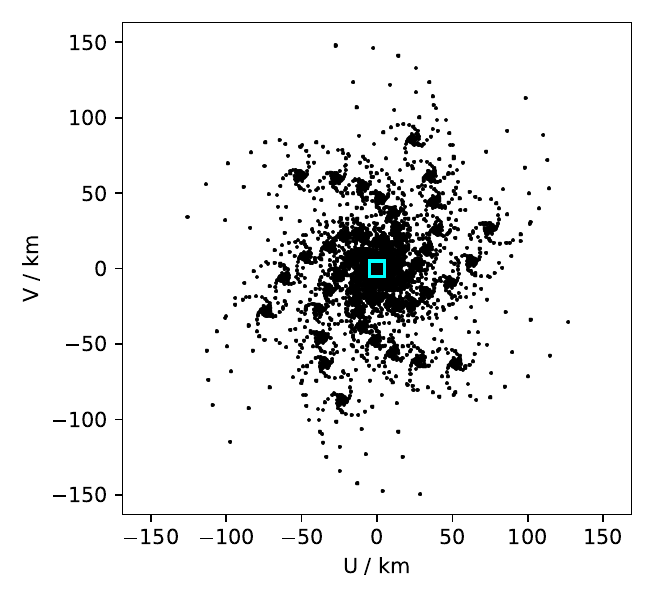}\,
\includegraphics[width=0.53\textwidth]{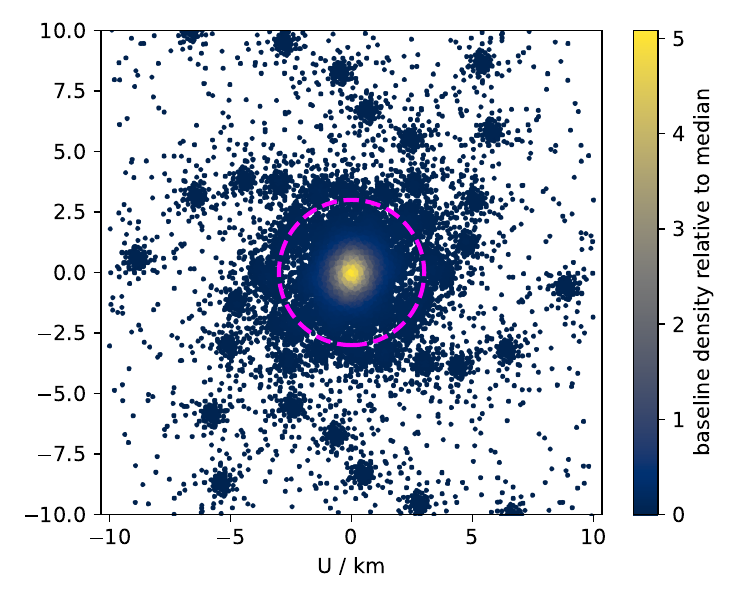} \\
\includegraphics[width=0.83\textwidth]{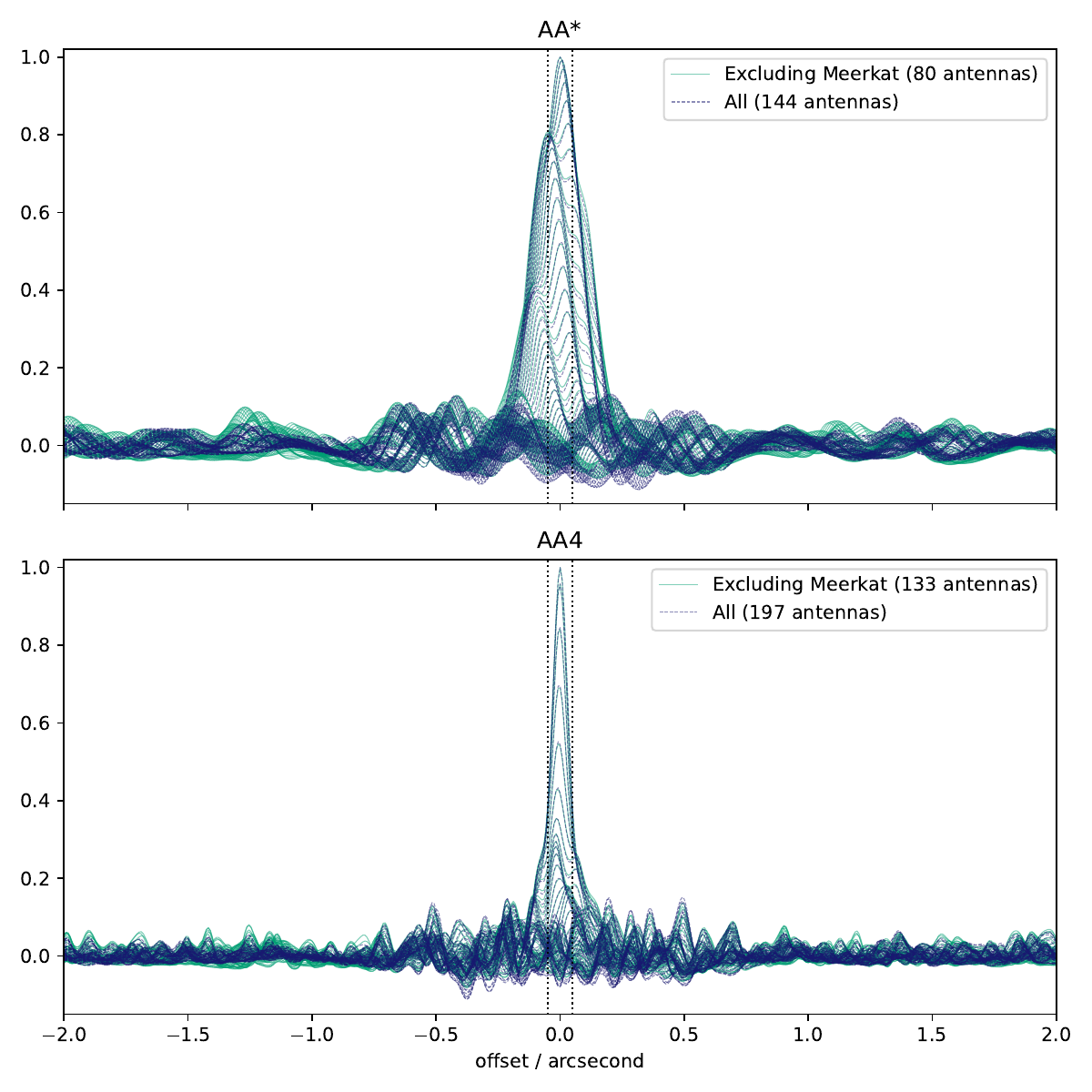}
\caption{Top left: Instantaneous monochromatic UV coverage of SKA-Mid AA4 observing the Sun at midday in December. Top Right: Central area of the UV plane (indicated by cyan square in left panel). Magenta circle indicates a 3\,km radius. Color scale indicates density of baselines in the UV plane. Center and Bottom: Instantaneous monochromatic PSF at 10\,GHz (uniform weighting). Both AA* and AA4 configurations are shown, with both full array and excluding Meerkat dishes. Dotted vertical lines are separated by 0.1\arcsec.}\label{fig:psf}
\end{figure}

When SKA-Mid construction reaches the Array Assembly 4 (AA4) milestone, it would become the only radio astronomy observatory of capturing (sub)arcsecond, high-fidelity images in its frequency range. For example, at $10$\,GHz, SKA-Mid AA4 observing the Sun in December would have an instantaneous beam size of 0.1\arcsec, while the AA* subarray would have a beam size of 0.2\arcsec\ (see Figure~\ref{fig:psf}).
However, it should be noted that resolution across the SKA-Mid band is likely to be limited by scatter broadening in the intervening corona rather than the resolution of the instrument in the typical case.
This important aspect is discussed further in Section~\ref{subsec:scatter}.

At frequencies available to SKA-Mid, free-free emission from magnetized plasma in the corona leads to circular polarization signals that are a probe of the coronal magnetic field~\citep[e.g., see observations performed with the Nobeyama Radioheliograph;][]{Iwai2013}.  Circular polarization signals from free-free emission arise because the ordinary (O) and extraordinary (X) modes have different optical depths in an atmosphere with a temperature gradient. For a review of the relevant radio emission and propagation mechanisms in the solar atmosphere, the reader is referred to~\citet{Gary:2023}. A summary of open problems addressable with microwave observations of the solar corona is discussed in~\citet{Chen:2023}.

To illustrate the diagnostic potential of SKA-Mid, we synthesised radio images from snapshots of a magnetohydrodynamics (MHD) simulation of a solar flare (Rempel et al., in prep; Martinez-Sykora et al., in prep). The single-fluid MHD model that takes into account magnetoconvection, self-consistent coronal heating, radiative transfer and field-aligned thermal conduction along magnetic field lines. Details of the model and simulation setups are described in~\citet[][]{Cheung:2019} and~\citet{Rempel:2023}. These simulations were developed to guide the development of science and technical requirements for NASA's upcoming Medium Class Explorer mission, the Multi-slit Solar Explorer~\citep[MUSE;][]{DePontieu:MUSE,DePontieu:MUSE2022,Cheung:MUSE2022}. MUSE will observe the solar corona at sub-arcsecond resolution, delivering active-region scale spectral rasters at cadences of 10-20s to capture the spatiotemporal evolution of the solar corona. Additionally, the Solar-C/EUVST mission~\citep{SolarC} would provide subarcsecond resolution UV/EUV spectra covering spectral lines that span the full temperature range (photosphere to flare corona) of the solar atmosphere. The two missions are scheduled to launch before science verification for SKA-Mid is completed. Coordinated simultaneous observations with ground-based solar optical/infrared telescopes with similar local sidereal time as SKA-Mid, including the Swedish Solar Telescope~\citep[][]{Scharmer:2003, Scharmer_2008, Lofdahl:2021} and GREGOR~\citep{Kleint:Gregor} in the Canary Islands, would provide complementary observations of the photosphere and chromosphere at sub-arcsecond resolution. The operational opportunities and challenges of coordinated observations are discussed in Section~\ref{sec:almalessons}.

\begin{figure}
\centering
\includegraphics[width=0.9\textwidth]{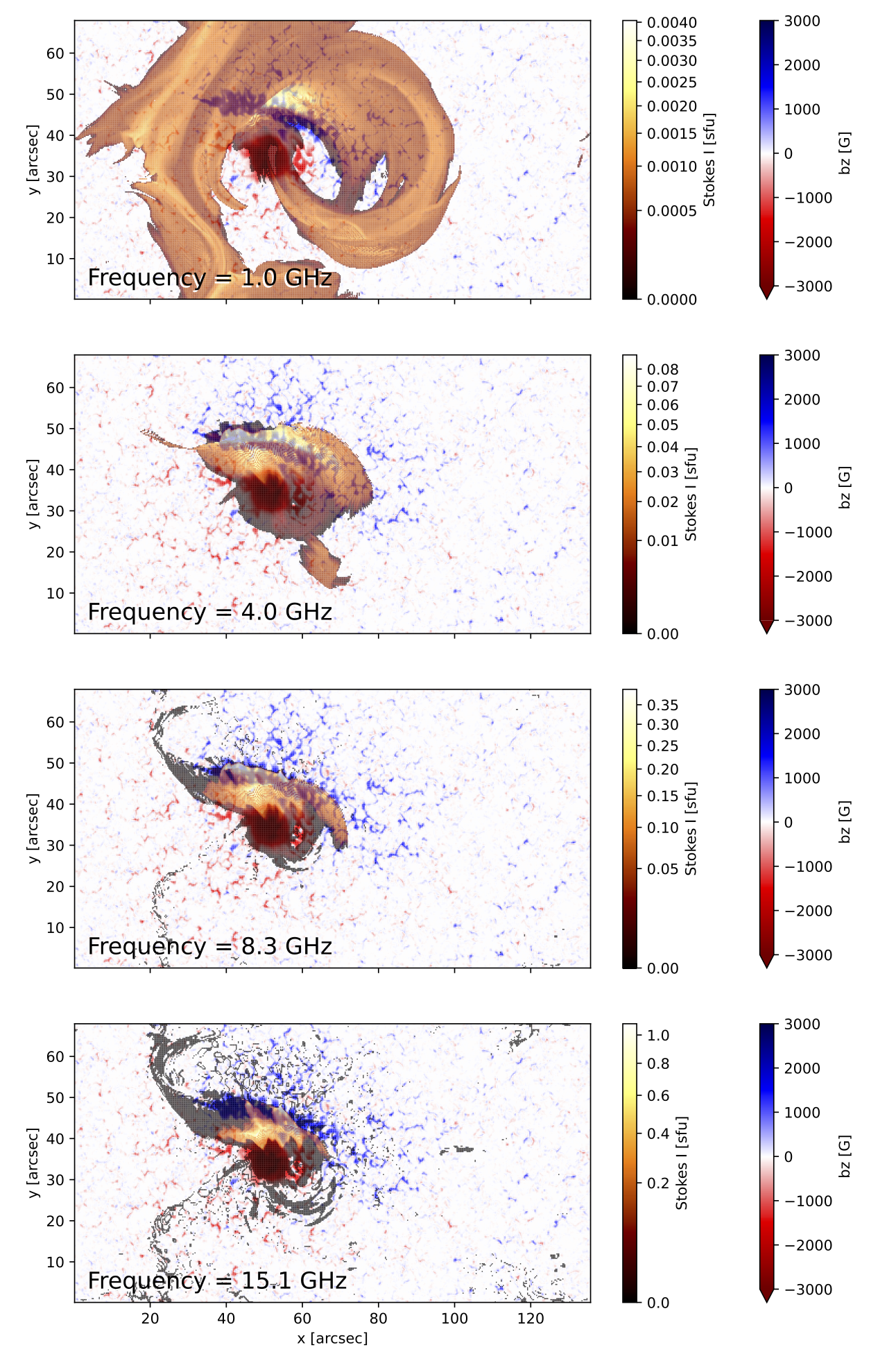}
\caption{Synthetic radio images of a simulated solar flare at $\nu = 1.0,4.0,8.3$ and $15.1$\,GHz. Each of the radio images is overlaid above the photospheric magnetogram showing the distribution of a pair of opposite polarity sunspots that are undergoing shearing motion.}\label{fig:multif_panels}
\end{figure}

Following~\citet{Rempel:2023}, this simulation models a flare by driving two opposing photospheric polarity regions (sunspots) against each other, leading to the formation of a flux rope which eventually results in an eruptive flare. The numerical grid spacing of the simulation is $192$\,km in the horizontal directions, and $64$\,km in the vertical direction, corresponding to 0.26\arcsec\ and 0.088\arcsec\ respectively. Figure~\ref{fig:multif_panels} shows the synthesized radio emission over the simulation domain for lines-of-sight parallel to the vertical axis of the simulation (i.e. top-down view). The synthesis was performed by extracting columns of the plasma (electron) temperature, electron density and magnetic field vector sampled along the line of sight, and synthesising the free-free and gyroresonance emission for the multi-thermal plasma populations using the method and code detailed in~\citet{Fleishman:GRFF}. The code can compute gyro-emission in various energy regimes and for different electron distributions (including kappa distributions), so comparisons of forward-modeled spectra and SKA-Mid observations would enable ruling out certain physics models of coronal heating and particle acceleration. 

The radio emission maps in Figure~\ref{fig:multif_panels} are displayed for frequencies that fall in the SKA-Mid band, 1, 4, 5a and 5b receivers. We note that the band~4 receiver covering 2.8 to 5.18\,GHz is part of the baseline design but not one of the first generation receivers. For $ \nu \lesssim 10$ GHz, the emission is dominated by gyroresonance (as opposed to free-free). The image at 1\,GHz shows loops associated with the erupting flux rope, whereas images at higher frequencies progressively show magnetic loops closer to the polarity inversion line. SKA-Mid observations would provide observational constraints to test models of coronal flare plasma that consist predominantly of multi-thermal plasmas~\citep{Cheung:2019} and models that consist of non-thermal populations~\citep[e.g.][]{Krucker:2014, Chen:2023, Chen:2024}. 

\begin{figure}
\includegraphics[]{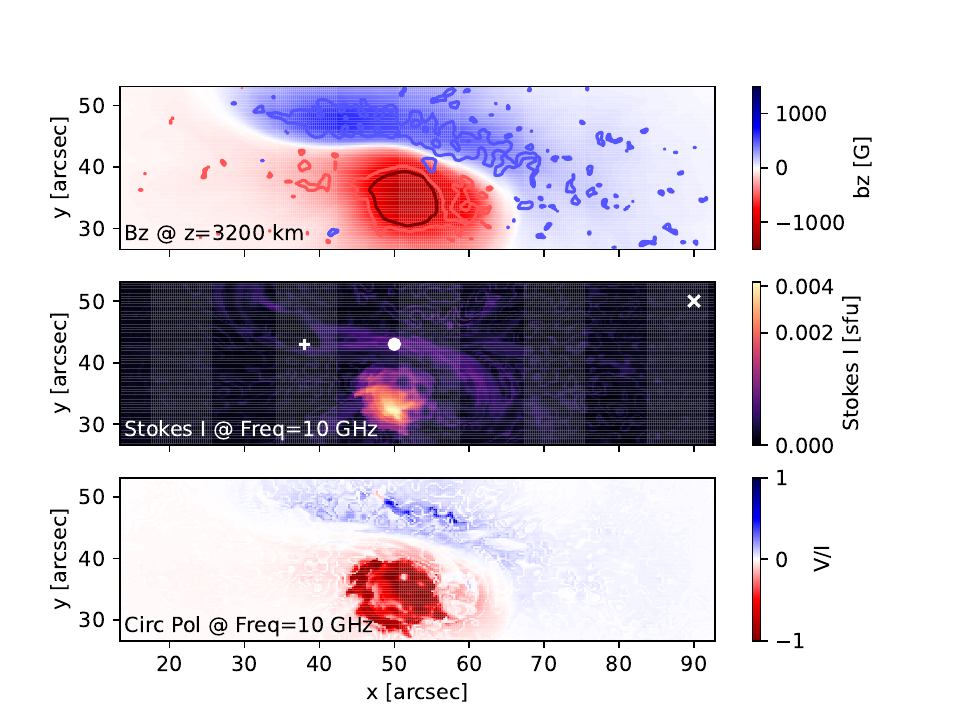}
\caption{Preflare state from a MURaM radiative MHD simulation of an M-flare (Rempel et al., in prep). Top panel: Distribution of line of sight magnetic field at the photosphere (contours at $\pm1,2$ kG; red is negative, blue positive) and at a height of 3200\,km (i.e. coronal field). Middle panel: Synthesized Stokes I at 10\,GHz. Bottom panel: Degree of circular polarization at 10\,GHz.} \label{fig:preflare}
\end{figure}

\begin{figure}
\includegraphics[]{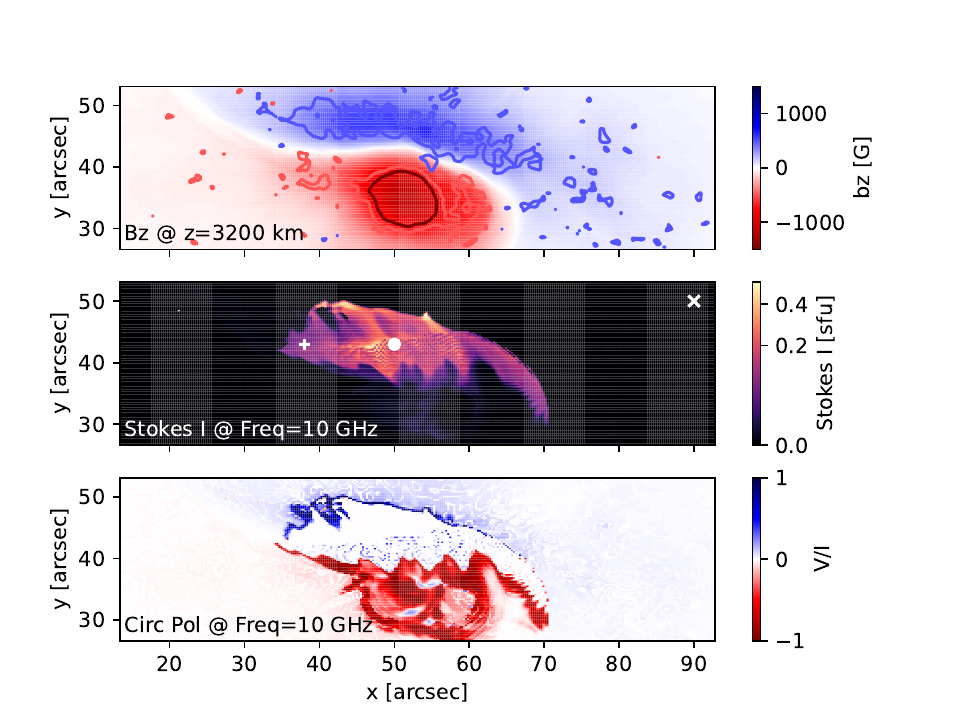}
\caption{Same as Figure~\ref{fig:preflare} but 580s later during the flare. See Figure~\ref{fig:multif_panels} for the full field-of-view of the model for this snapshot. The synthetic spectra sampled at the locations indicated by the $+$, $\bullet$ and $\times$ markers in the middle panel are displayed in Fig.~\ref{fig:spectra}.}\label{fig:flare}
\end{figure}

Figure~\ref{fig:preflare} shows a comparison between synthetic magnetograms at the photosphere and 3200\,km above (top panel), as well as the map of circular polarization at $\nu=10$\,GHz. This preflare snapshot precedes the flare snapshot (Figure~\ref{fig:flare}; same time as Figure~\ref{fig:multif_panels}) by 580s. The magnetograms and circular polarization maps show morphological correspondence, consistent with previous microwave observations~\citep[e.g. from Nobeyama Radioheliograph, see][]{Iwai2013}. In the flare state, the corrugation of edge of the flare loop system are apparent in the Stokes I and circular polarization images and could be a sign of MHD instabilities flare current sheet outflow~\citep[][]{French:2021}. Such signatures would be difficult to detect without observations resolved at the arcsecond scale. 

\begin{figure}
\includegraphics[width=\textwidth]{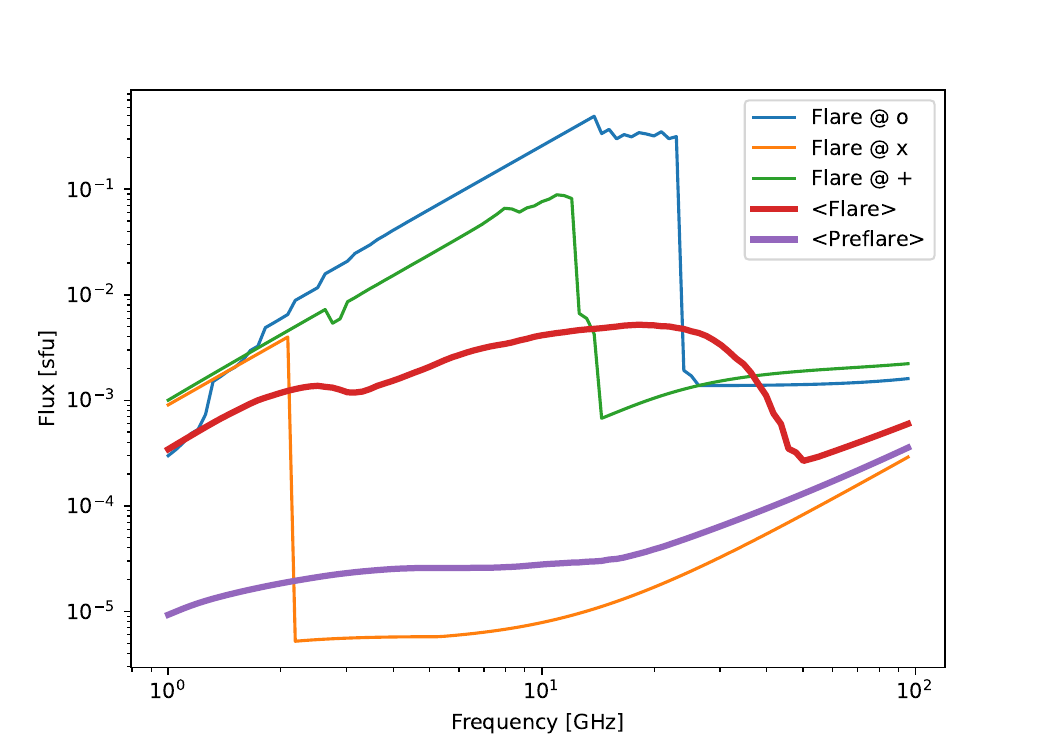}
\caption{Synthetic radio emission spectra from the radiative MHD simulation of a solar flare. Spectra sampled at the locations indicated by the $+$, $\bullet$ and $\times$ markers in Figure~\ref{fig:flare} are displayed. Overplotted are the area-averaged spectra over the entire computational domain (full extent shown in Figure~\ref{fig:multif_panels}) for the preflare (Figure~\ref{fig:preflare}) and flare (Figure~\ref{fig:flare}) snapshots. The preflare emission is dominated by gyroresonance at the frequencies below roughly 10\, GHz, and by free-free emission at higher frequencies. }\label{fig:spectra}
\end{figure}


High-resolution imaging observations with SKA-Mid will provide important constraints on the fundamental physics of solar flares.  To illustrate this point, Figure~\ref{fig:spectra} shows the synthetic spectra spatially-averaged over the entire computational domain of the preflare (Figure~\ref{fig:preflare}) and flare (Figure~\ref{fig:flare}) snapshots. The spatially-averaged preflare spectrum is dominated by gyroresonance emission below roughly 10\,GHz and by free-free emission at higher frequencies, where it monotonically increases with frequency. In comparison, the spatially-averaged flare spectrum peaks at $\sim$20\,GHz (for this particular model) and is dominated by gyroresonance emission. Also plotted in this figure are spectra sampled at specific locations in the flare state (see the $+$, $\bullet$ and $\times$ markers in Figure~\ref{fig:flare}). The spectra sampled at these locations are also peaked, but unlike the spatially-averaged spectrum, they have a much steeper drop with frequency post-peak, as is characteristic of \emph{iso}-thermal plasmas exhibiting gyroresonance emission. Due to the existence of multi-thermal plasmas in simulated flare, spatial averaging tends to smooth out this precipitous drop, making the spectrum appear less steep and less thermal-like. Spatially-resolved radio spectra would test when single-fluid MHD models like this one are no longer valid, and where/when non-thermal particle acceleration effects are important. Such observations would also constrain the coronal heating \citep[see, e.g.,][]{2025ApJ...988..100F} and the heating mechanisms at flare loop footpoints important for flare manifestations at other wavelengths~\citep[e.g., see][]{Cheung:MUSE2022,Kerr:2024}.

\section{Synthetic SKA-Mid Observations}

As discussed in Section~\ref{sec:SKAmidSolar}, SKA-Mid AA4 offers dense UV coverage as well as long baselines, a combination which promises to provide high-fidelity and high resolution imaging.
However, solar observations are particularly challenging for multiple reasons including:
(1) there is diffuse emission at multiple length-scales, and
(2) the changing solar conditions do not permit rotation synthesis to be used to improve UV coverage (this also applies to some extent to multi-frequency synthesis, since emission also evolves with frequency).
In order to evaluate the ability to reconstruct solar images, we used the Karabo software package~\citep{Karabo:2025} to simulate an observation for a sky model corresponding to the scene in Figure~\ref{fig:preflare}.

\subsection{Generation of synthetic visibilities}
Karabo takes a sky model in right ascension (RA) and declination (DEC) coordinates and flux densities in Jansky.
We converted model coordinates into a Karabo input map with RA and \postreview{DEC} coordinates, with the center of the adjusted field of view to point at the zenith for SKA-Mid. We evaluated the image quality due to dense SKA UV coverage at two representative frequencies: 5\,GHz and 10\,GHz.
All AA4 antennas are used, although excluding MeerKAT~\citep{MeerKAT:2016} dishes makes very little difference (see Figure~\ref{fig:psf}). Input maps from the MHD simulation at these two frequencies are processed independently.
Conservatively, only a narrow frequency coverage of 10\,kHz was used; not nearly enough to increase UV coverage via multi-frequency synthesis.
The current treatment does not include other effects such as system noise and source scattering (see Section~\ref{subsec:scatter} for a further discussion).

\begin{figure}
\includegraphics[width=\textwidth]{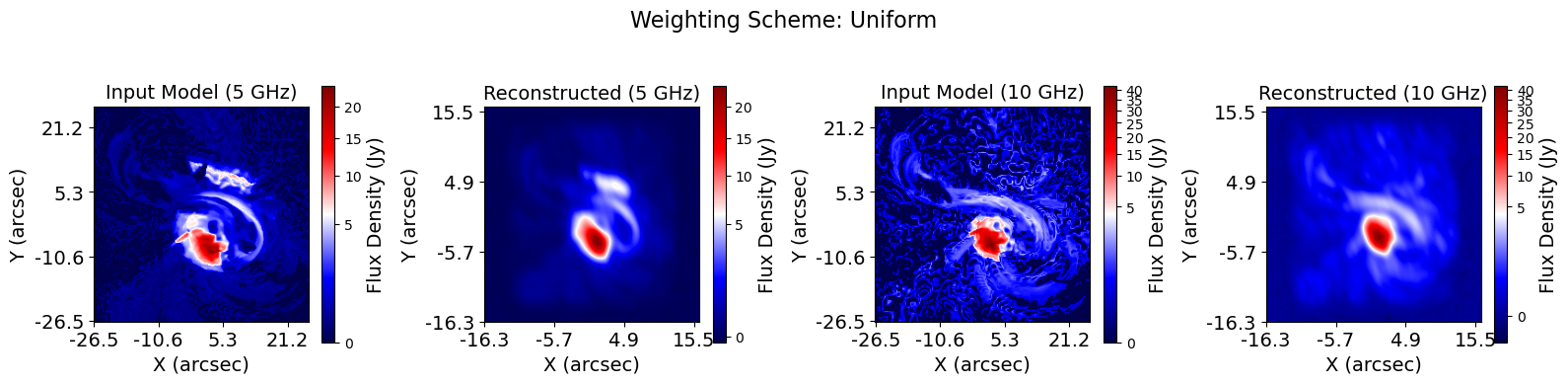}
\includegraphics[width=\textwidth]{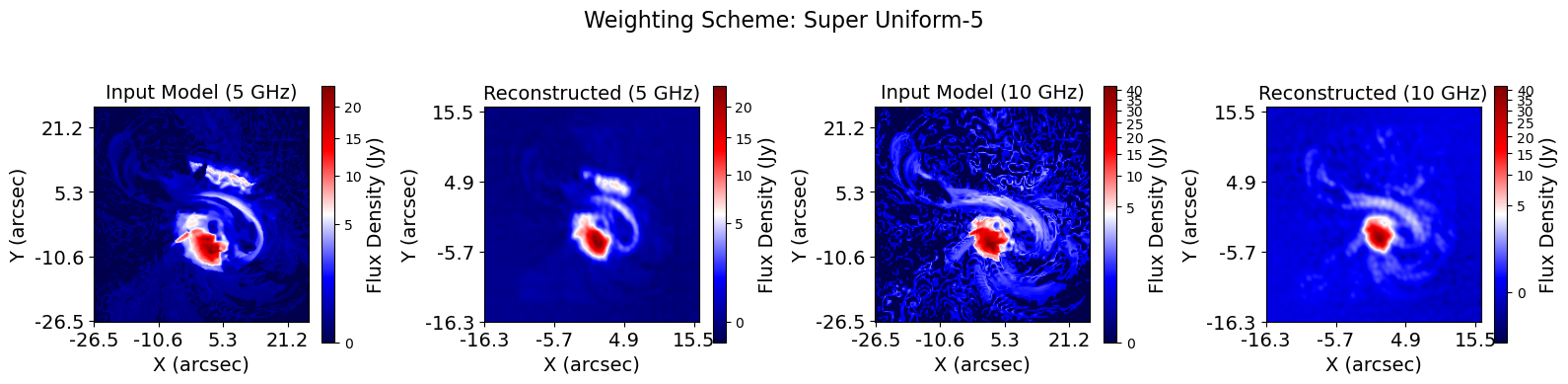} \caption{Radio emission images at $5$\,GHz and $10$\,GHz synthesized from the MHD model, and image reconstructions using different weighting schemes: uniform (top) and super-uniform (bottom).}\label{fig:reconstruction}
\end{figure}

\subsection{Imaging}
Karabo outputs a measurement set of the mock observation, which is used for deconvolution with WSClean \citep{2014MNRAS.444..606O}.
We chose standard imaging parameters with pixel size (0.05\arcsec) as approximately $1/5$th of the synthesised beam, and total image size $\sim200\arcsec$.
Since the rich morphology of the solar flare emission must be captured accurately, the weighting scheme to be used is crucial.
We did experiments with different weighting schemes, namely natural, uniform, and super-uniform.
We find that the natural scheme is an extremely poor choice for SKA-Mid, due to the very high density of intra-core baselines.
As shown in Figure~\ref{fig:psf}, SKA-Mid has exactly half of its antennas in a circle of radius $500$\,m in the core array.
This makes a more uniform weighting scheme vital in order to avoid a wide pedestal in the synthesized beam.
For solar imaging, this does not present a problem, since the emission we are interested in is extremely bright, and so sensitivity considerations, which might motivate a more natural weighting scheme, do not apply.
Instead, we wish to maximize the resolution and (arguably more importantly) minimize sidelobes of the PSF.
Images based on Uniform and Super-uniform weighting schemes, on the other hand, are capable of recovering the detailed structure of the flaring loops. Figure~\ref{fig:reconstruction} shows the input flare sky model and the reconstructed maps for 5\,GHz and 10\,GHz central frequencies. \postreview{In Figure~\ref{fig:reconstruction}, we note that the typical dynamic range of the simulated images is $\approx 230$ and $\approx 160$, with peak flux densities of the source at $23$ SFU at 5 GHz and $43$ SFU at 10 GHz, respectively. Therefore, we expect to detect fine structures down to $\approx 0.1$ SFU at 5 GHz.}

\postreview{The SKA-Mid receiver systems feature minimal cross-polarization coupling. In fact, components inside the modular cryostat are built to strict specifications, limiting intrinsic hardware leakage to less than 1\% in power \citep{2013JASS...30..345C, 2019MNRAS.488.1618B}. However, the actual level of leakage changes drastically depending on where it is measured: the raw hardware itself, the wide-field effects across the dish beam, or the final data after software calibration. Therefore, for a fully circularly polarised flare source, with peak flux density $\approx23$ SFU at 5 GHz (Fig. \ref{fig:reconstruction}), one expects to detect sub-SFU flux density polarised emission under ideal scenarios. The detection limits for polarisation will scale with intrinsic circular polarisation levels, propagation effects and leakage noise from all dishes. The impact of these effects on the images is under investigation with the Karabo pipeline.}

We reiterate that we have used only a 10\,kHz bandwidth in order to produce these maps.
A wider fractional bandwidth would increase UV coverage, and given the wideband nature of the radio emission we are probing, this is a possible path to increasing image fidelity further, however care would need to be taken, both in the impact of this assumption, and in handling the effect of the primary beam of the dishes, which also varies with frequency.

\subsection{Scattering and angular broadening}
\label{subsec:scatter}
The synthetic observations presented here do not take into account potential scattering and refraction effects. As discussed in~\citet[][see also \citeauthor{Bastian:1994},~\citeyear{Bastian:1994}]{Gary:2023}, angular broadening due to scattering in the intervening magneto-ionic medium of the corona is approximately 20\arcsec/$\left(\nu/\mathrm{GHz}\right)$; or, equivalently, the resolution of a 3-km baseline. This would allow subarcsecond imaging under typical scattering conditions only at frequencies above 5\,GHz.

We note that 3\,km is almost exactly the median baseline length of AA4\footnote{In contrast to the AA4 array, the AA* subarray has much sparser coverage of these intermediate baseline lengths, resulting in a less Gaussian central lobe of the PSF (see Figure~\ref{fig:psf}). AA4 may be a considerable improvement for high-fidelity imaging, especially in the snapshot, narrowband case considered here.} of SKA-Mid, and so the array is extremely well suited to observations made under typical observing conditions.
This is illustrated in the top-right panel of Figure~\ref{fig:psf}, which shows the baselines out to 10\,km.
The UV coverage out to 3\,km is exceptionally good, although uniform weighting is still necessary due to the very large number of short, intra-core baselines.

However, at least initially, it will be highly desirable to make solar observations with the full SKA.
This is because line-of-sight scattering is likely to be highly variable, and so the full range of SKA baselines will be required firstly to exploit lower-than-average scattering conditions.
Additionally, measurements of scattering are interesting in themselves, and so SKA-Mid observations will provide stringent constraints on the theory. \postacceptance{This is discussed further in \cite{Dey01.2026.SKA}}.

\section{Operational considerations}

Solar flares are classified by their peak X-ray flux in the 1-8 \AA~band. The simulation presented here is an M-class flare, corresponding to one with a peak flux above $10^{-5}$ Wm$^{-2}$. The strongest flares (X-class) would have fluxes exceeding $10^{-4}$ Wm$^{-2}$. During solar cycle maximum, M-class flares are quite frequent, numbering in the hundreds per year and diminishing to practically non-existent during solar minimum. The occurrence of a flare from a sunspot group in the day prior is associated with an increased probability for a flare the following day~\citep{Park:2020}. With appropriate planning and flare forecasting, it is very achievable to capture flares of this size using SKA-Mid. There is a continuous distribution of flares with lower peak X-ray fluxes and increased frequency, down to the scale of nanoflares, which are associated with heating events that sustain the million-Kelvin corona. Though not the focus of this paper, SKA-Mid observations of smaller flares and even the quiet Sun would be highly valuable. \postacceptance{\cite{Mondal01.2026.SKA} discusses this aspect of detecting quiet Sun non-thermal emission using SKA-low and SKA-mid observations.}

The solar disk subtends an angle of approximately 0.5$^\circ$, and so at the higher SKA-Mid frequencies it will not be able to cover the whole of the solar disk in a single pointing. However, it will still be possible to capture many bright flares by tracking the most promising sunspot group. This will also ensure the most accurate and reliable polarimetry by keeping the highly polarized emission close to boresight. 

It is noteworthy that solar observations have been successfully undertaken with MeerKAT. \postreview{\citet{2025FrASS..1266743K} successfully demonstrated whole-Sun imaging in the UHF (580–1015 MHz) and L-band (900–1670 MHz). \citet{Luo:2026} demonstrated imaging spectroscopy of an M-class solar flare, achieving a dynamic range exceeding 10$^3$. The observing and data processing techniques of both studies should be extensible to SKA-Mid.} \postacceptance{We refer the reader to~\citet{Oberoi01.2026.SKA} for a detailed discussion of calibration strategies to achieve high dynamic range and polarimetric sensitivity for SKAO.}

\subsection{Lessons learned from solar observing with ALMA}
\label{sec:almalessons}
Observing the Sun with radio interferometers presents substantial operational challenges that differ in important ways from observing other astronomical targets. 
In this respect, important lessons can be learned from the development of solar observing and data processing strategies for the Atacama Large Millimeter/submillimeter Array 
\citep[ALMA,][]{2018Msngr.171...25B,2017SoPh..292...87S,2017SoPh..292...88W,2016SSRv..200....1W}. 
While ALMA is observing the Sun at higher frequencies (currently about 92--355\,GHz) than SKA, similar challenges can be expected. 

\paragraph{Snapshot Imaging.} 
The solar atmosphere evolves on timescales of seconds to minutes across a wide range of spatial scales. This makes long integrations, as commonly used to fill the uv-plane for other astronomical targets, unsuitable. 
ALMA's solar observing modes therefore rely on snapshot imaging with very short integration times to freeze the dynamics. 
Despite numerous baselines from up to 62 antennas, ALMA's resulting instantaneous $uv$-coverage remains sparse, which leads to degraded image fidelity compared to a telescope with a filled aperture of equivalent size. 
The large number of antennas for SKA-Mid improves the situation but nonetheless cannot completely remove the challenges with imaging based on marginal uv-sampling.
This is exacerbated by the high concentration of antennas in the core for SKA-Mid (with exactly half the antennas within a central circle of radius 500\,m).
The longer baselines of SKA-Mid will further emphasise the conflict between temporal resolution and spatial sensitivity. Solar-tailored observing modes (e.g. fast scanning strategies and novel uv-interpolation methods) will be required to mitigate sparse sampling while ensure that accurate mapping of rapidly evolving solar structures, which is crucial to most solar science cases. 

\paragraph{Data processing} Processing interferometric solar data and reconstructing high-quality maps of the observed regions is a highly complicated task, particularly in view of unavoidable instrumental noise, atmospheric degradation, and incomplete uv-coverage.
For decades, the CLEAN algorithm in its various implementations has been the standard approach for radio interferometric imaging. However, it is suboptimal for extended sources such as the Sun, which fill the primary beam with highly structured and spatially complex emission. It is therefore advisable to explore alternative imaging strategies for solar observations with the SKA, such as the Maximum Entropy Method 
\citep[MEM,][]{1985A&A...143...77C}, the Adaptive Scale Pixel \citep[ASP,][]{2004A&A...426..747B} deconvolution algorithm, or emerging machine-learning-assisted techniques.

A fundamental challenge remains the absence of a true reference (``ground truth'') against which the performance of reconstruction methods can be quantitatively assessed and optimized. The SolarALMASimulator \citep[SASim,][]{2024arXiv240814265W}, developed for simulating ALMA observations of the Sun, demonstrates the potential of forward modelling of the solar atmosphere, Earth's atmosphere, and the instrumental response.
A similar approach for SKA-Mid would offer substantial potential for optimizing both observing strategies and data-processing pipelines for solar interferometric observations, but would require a dedicated effort. The continued development and refinement of modeling packages like Karabo~\citep{Karabo:2025} is encouraged.

 
\paragraph{Coordination and scheduling constraints.}
Solar observations are often carried out as coordinated campaigns that involve multiple space-borne and ground-based observatories. 
While this makes the operations more challenging, this strategy is motivated by the need to combine complementary diagnostics that sample different parts and properties of the solar atmosphere. 
Such complementary data sets are necessary in order to unveil the physical processes behind the complex phenomena observed on the Sun. 
The consequence is that all involved observatories must carry out their scientific observations in the exact same time window, which is not common for other astronomical observations. 
Deviations from the agreed schedule, e.g., due to local weather or technical delays, must be avoided in order to maximize the scientific value of such coordinated campaigns. 
Successful execution requires awareness and good communication between all involved parties. 

SKA-Mid's geographical location is an important factor in this context. The small time difference to solar telescopes on the Canary Islands, including the planned 4\,m European Solar Telescope (EST), facilitates coordinated observations. In contrast, joint observations with more distant facilities are constrained by larger longitudinal separations. In particular, co-observing with the Daniel K. Inouye Solar Telescope (DKIST) on Hawaii is practically impossible, since the two observatories are on opposite sides of the Earth.

\paragraph{Acknowledgements}
This material is based upon work supported by the NSF National Center for Atmospheric Research, which is a major facility sponsored by the U.S. National Science Foundation under Cooperative Agreement No. 1852977. We would like to acknowledge high-performance computing support from the Derecho system (doi:10.5065/qx9a-pg09) provided by the NSF National Center for Atmospheric Research (NCAR), sponsored by the National Science Foundation. 
SW acknowledges support by the Research Council of Norway, project number 325491, through its Centres of Excellence scheme, project number 262622 (RoCS). GF was supported in part by NSF grant  
AGS-2425102.

\bibliographystyle{abbrvnat-maxbibnames4}
\bibliography{chapter} 

\end{document}